\documentclass[letterpaper,aps,prl,preprint,groupedaddress,showpacs,amsmath,amssymb,endfloats*]{revtex4}
\usepackage{graphicx}
\usepackage{subfigure}
\usepackage{dcolumn}
\begin{document}

\title{Temperature Effects on Threshold Counterion Concentration to Induce
Aggregation of fd Virus}

\author{Qi Wen}
\affiliation{Department of Physics, Brown University, Providence, RI 02912 }
\author{Jay X Tang }
\email[Email to: ]{jxtang@physics.brown.edu}
\affiliation{Department of Physics, Brown University, Providence, RI 02912 }

\date{\today}

\begin{abstract}
We seek to determine the mechanism of like-charge
attraction by measuring the temperature dependence of critical
divalent counterion concentration ($\rm{C_{c}}$) for the
aggregation of fd viruses. We find that an increase in temperature causes $\rm{C_c}$ to decrease, primarily due to a decrease 
in the dielectric constant ($\varepsilon$) of the solvent. 
At a constant $\varepsilon$, $\rm{C_c}$ is found to increase as the temperature increases. The effects of $T$ and $\varepsilon$ on $\rm {C_{c}}$ 
can be combined to that of one parameter: Bjerrum length ($l_{B}$). $\rm{
C_{c}}$ decreases exponentially as $l_{B}$ increases, suggesting that entropic effect
of counterions plays an important role at the onset of bundle
formation. 
\end{abstract}


\maketitle
Multivalent counterion induced aggregation of polyelectrolytes has
been observed in a variety of systems such as DNA, F-actin and bacteriophages~\cite{Jay_JBC1996,Butler_PRL2003,Raspaund_BJ1998,Bloomfield_BP1995}.
During the past decades, there have been intensive theoretical investigations
to elucidate the mechanism of counterion induced attractive interaction
between polyelectrolytes~\cite{Ha_PRL1997,Jensen_PRL1997,shklovskii_PRL99,Grosberg_RMP2002,Nguyen_JCP2000,Solis_PRE1999,Levin_EPB1999,Parsegian_BJ1992,Rau_Biochem2004}.
Although other types of interactions such as hydration~\cite{Parsegian_BJ1992,Rau_Biochem2004}
and depletion~\cite{Tang_biochem1997} are contributing factors, it has been established
that the electrostatic between the polyelectrolytes and their correlated counterions~\cite{shklovskii_PRL99,Nguyen_JCP2000,Jensen_PRL1997,Ha_PRL1997,Ha_PRL1998,Solis_PRE1999,Grosberg_RMP2002,Levin_EPB1999}
is the major cause of like-charge attraction. 

There are, however, two possible ways, in which the condensed counterions on 
different polyelectrolytes correlate with each other. 
Thermal fluctuations create transient regions of high or low 
counterion density along the polyelectrolytes, which are typically simplified to be charged lines each with a thin layer of 
condensed counterions~\cite{Oosawa_PE1968}. When two parallel polyelectrolytes get close to each 
other, an attractive force is induced by their transient complementary counterion density profiles. The 
attractive force induced by the long wavelength counterion fluctuations has been predicted to be proportional to temperature using the mean field 
theory~\cite{Oosawa_PE1968}. Alternatively, the counterions may correlate with each other in their positions 
on the surfaces of polyelectrolytes. A representative picture of the
attractive interaction induced by positional correlations of counterions
is provided by the Wigner crystal model~\cite{shklovskii_PRL99}.
In this model, condensed counterions form Wigner crystals on the polyelectrolyte's
surface at $T=0$ K. Cross correlation of counterions occurs when the distance between two polyelectrolytes decreases to the lattice constant of the Wigner crystals. It is the cohesive energy of the Wigner crystals
that results in attractive interaction and aggregation of like-charged
rods~\cite{Rouzina_JPC1996,shklovskii_PRL99}. Since thermal fluctuation
diminishes the structural order of counterions, this model predicts
a stronger attractive interaction at a lower temperature~\cite{Jensen_PRL1997}.

Recent computer simulation results suggest that the Wigner crystal
model captures the physics of like-charge attraction better than
the thermal fluctuation model~\cite{Jensen_PRL1997}, since it takes
into account the strong correlations between individual
counterions. However, the Wigner crystal model is a zero
temperature approximation. It is uncertain whether the model works under high temperatures, even though it
has been suggested in the theoretical work that the short range
order of counterions survives at high
temperature~\cite{Jensen_PRL1997,shklovskii_PRL99,Grosberg_RMP2002}.
Since the two theoretical models predict opposite temperature
effects on attractive interactions, the aggregation of
polyelectrolytes under different temperatures is expected to directly reveal the
dominant mechanism of like-charge attraction.

In this paper, we present the results of temperature effects on critical concentration of counterions ${\rm {C}_{c}}$,
using ${\rm {MgCl}_{2}}$ and ${\rm {CaCl}_{2}}$ to induce bundle formation
of bacteriophage fd. We found that ${\rm {\rm {C}_{c}}}$ increases
with increasing temperature under a certain dielectric constant. The
effect of the dielectric constant on ${\rm {C}_{c}}$ has been measured
as well. Under a fixed temperature, ${\rm {C}_{c}}$ decreases with
decreasing $\varepsilon$. Noting the similar effects of $T$ and
$\varepsilon$ on threshold concentration, we speculate that ${\rm {C_{c}}}$
is actually determined by the Bjerrum length ($l_{B}$). The result
that ${\rm {C_{c}}}$ doesn't change with temperature at a constant
$l_{B}$ validates our speculation. The thermodynamics for aggregation
of fd virus is investigated using the result that ${\rm {C_{c}}}$
drops exponentially as a function of $l_{B}$. The effect of hydration of 
counterions can be assessed from the prefactor in the exponential expression. 
Finally, based on the result that stronger counterion correlations
lead to lower threshold concentration, we conclude that our results support
the Wigner Crystal model.

Bacteriophage fd is a rod-like polyelectrolyte of approximately 880
nm in length, and 6.6 nm in diameter. There are approximately 2,700
copies of coat protein arranged crystallographically on the virus
surface. At neutral $p{\rm {H}}$, each coat protein contributes four net negative
charges on the surface of a virus. Thus, a virus has a linear charge density of 
approximately 12.5 e/nm. When a virus is simplified to a linear array of charges,
the charge spacing would be $b=0.8$ \AA. The fd viruses are highly charged so that 
their latteral aggregation can be induced by divalent
counterions such as $\rm{Mg}^{2+}$ and $\rm{Ca}^{2+}$~\cite{Jay_PRL98, jay_BJ2002}.
The viruses survive at temperatures
up to $90\phantom{.}^{\circ}{\rm {C}}$~\cite{Williams_BBA1984,Ji_Utramicroscopy}. Therefore, within 
our experimental range ($T\leq 50\phantom{.}^\circ\rm{C}$), the results are not expected to be due to the changes in the physical parameters of virus.

Aggregation of fd virus is detected by measuring the scattering light
intensity at a fixed angle of 90$^{\circ}$, with a PERKIN ELMER LS-5
luminescence spectrometer. 800 $\mu{\rm {l}}$ 0.1 mg/ml bacteriophage
fd was added to a rectangular cuvette of 10 mm path length and 5 mm width. Scattering
intensity was measured when the solution reached its steady state
following the addition of a stock solution of concentrated ${\rm {CaCl_{2}}}$
or ${\rm {MgCl_{2}}}$. Sample temperature was controlled using a ISO TEMP 1006S water bath (Fisher Sci Inc.)
connected to the sample holder. When increasing the counterion
concentration, an abrupt increase in the light scattering intensity
was noted at a critical concentration ${\rm {C_{c}}}$. The ${\rm {C_{c}}}$
was defined at the divalent concentration where at least a 20-fold
increase in the total scattering intensity was observed.

To find out the temperature effect on threshold concentration, we
first measured the threshold concentration under different temperatures
without other manipulation of solution properties. Next, the effect
of solution dielectric constant on ${\rm {C_{c}}}$ was measured under
a constant temperature of $20~^{\circ}{\rm {C}}$. Finally, ${\rm {C_{c}}}$
was measured at different temperatures with the dielectric constant fixed.

\begin{figure}[f]
\subfigure{\includegraphics[width=0.45\linewidth]{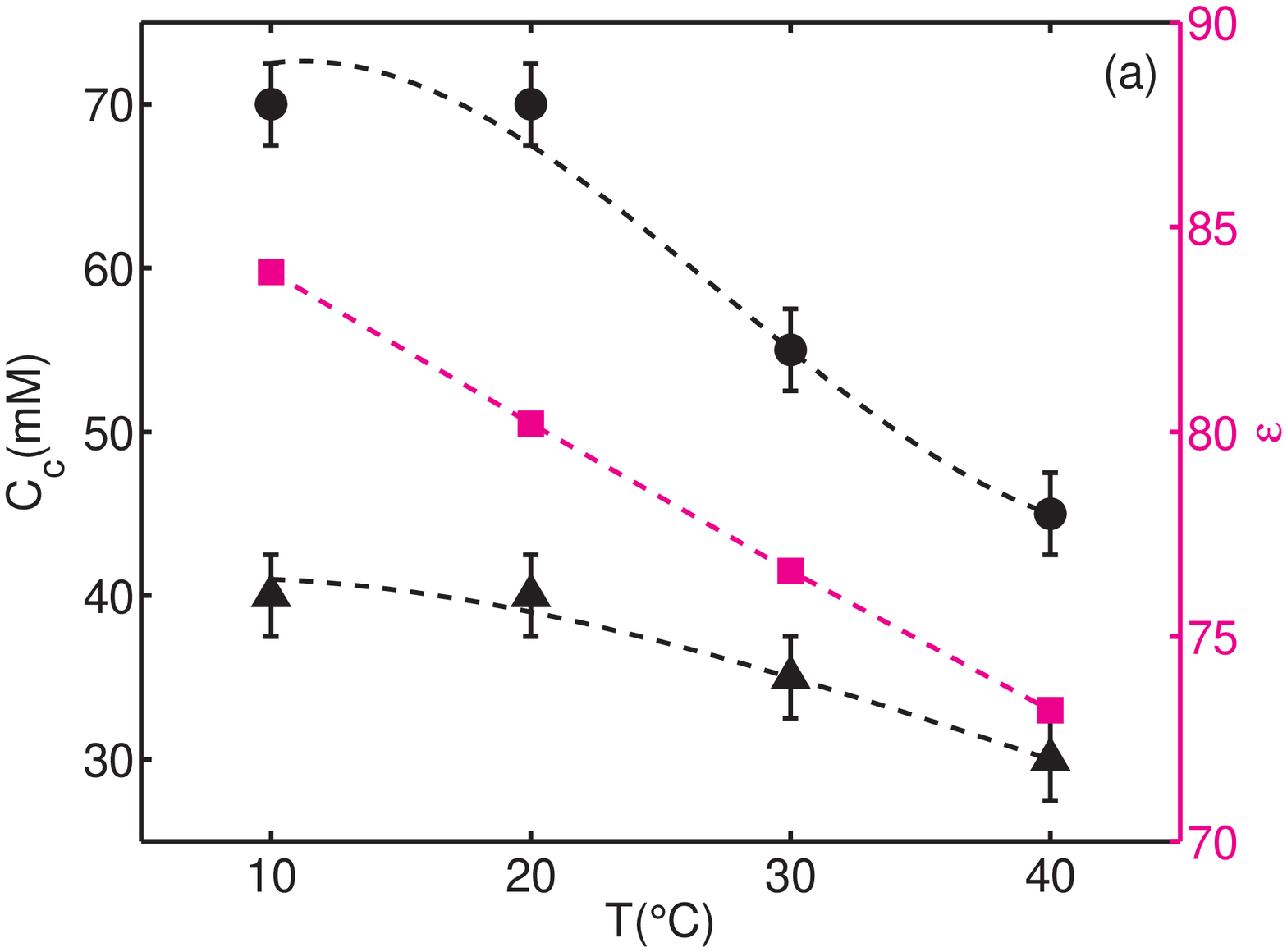}\label{fig1a}} \subfigure{\includegraphics[width=0.45\linewidth]{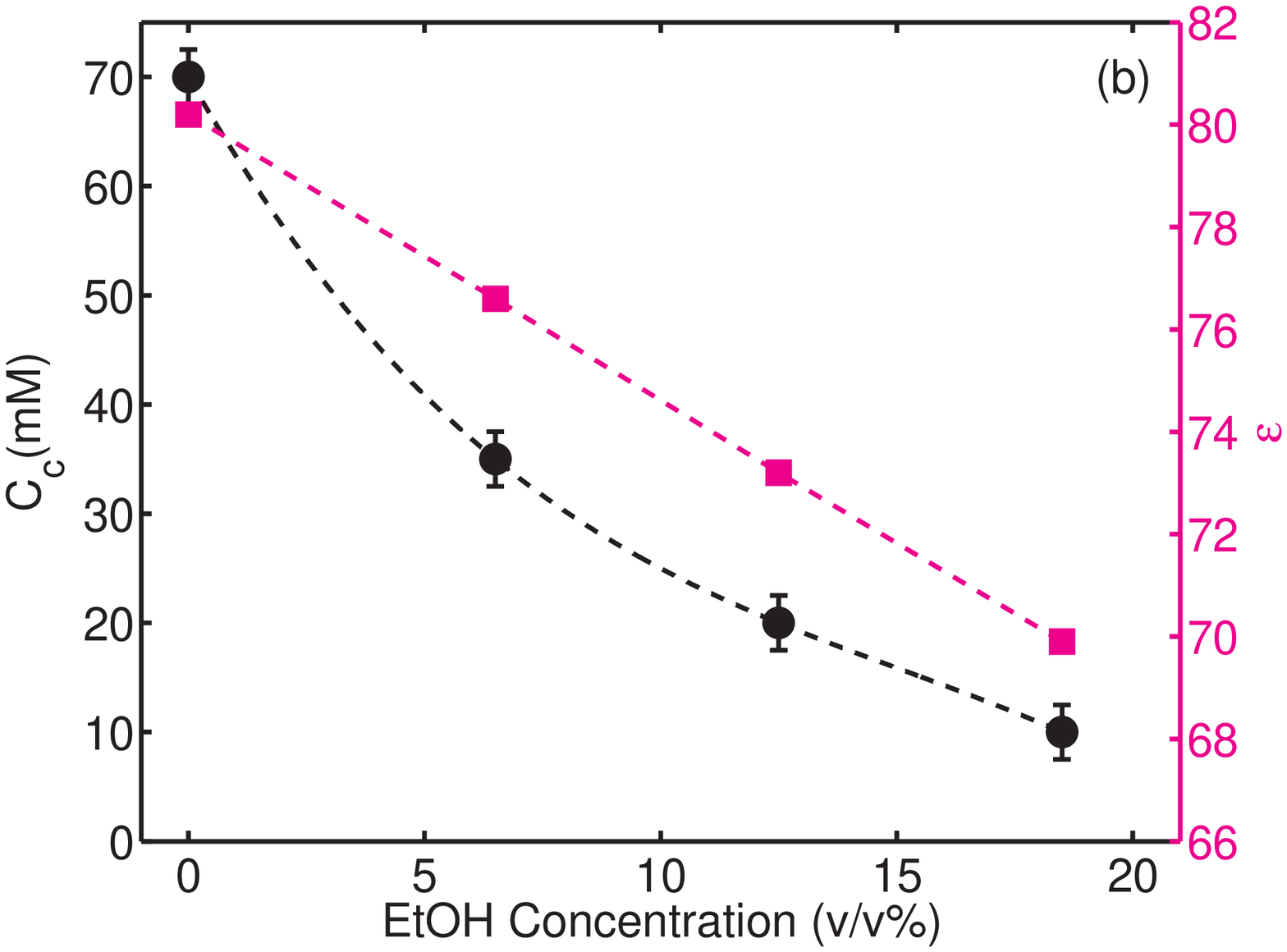}\label{fig1b}}
\caption{Increase of temperature leads to decrease of dielectric constant
($\varepsilon$) of the solvent, thus causes ${\rm {C_{c}}}$ to decrease.
(a) When the $\varepsilon$ of the solvent is not fixed, ${\rm {C_{c}}}$
of both ${\rm {MgCl_{2}}}$ (solid triangles) and ${\rm {CaCl_{2}}}$
(solid circles) decreases with increasing temperature. $\varepsilon$
(solid squares) drops as temperature increases. (b) ${\rm {C_{c}}}$ of ${\rm {MgCl_{2}}}$
(circles) decreases with the amount of ethanol at 20 $^{\circ}{\rm {C}}$. Addition
of ethanol leads to decrease of $\varepsilon$ (squares).\label{fig1}}
\end{figure}

Fig.\ref{fig1a} shows threshold concentrations of ${\rm {MgCl_{2}}}$
and ${\rm {CaCl_{2}}}$ as functions of the sample temperature. ${\rm {C_{c}}}$
for both ${\rm {MgCl_{2}}}$ and ${\rm {CaCl_{2}}}$ decreases with
increasing temperature. However, the temperature changes of a solution
also results in changes of its dielectric constant. The dielectric
constant of the solution, which is approximately equal to that of water,
under different temperatures are calculated using $\varepsilon(T)=a+bT+cT^{2}+dT^{3}$,
where $a$ , $b$ , $c$ , $d$ are empirical constants~\cite{CRCBOOK}.
It is shown in Fig.\ref{fig1a} that the calculated $\varepsilon$
of water decreases from 83.9 to 73.2 when temperature increases
from 10 to $40~^{\circ}{\rm {C}}$.

Changing the dielectric constant dramatically affects ${\rm {\rm {C_{c}}}}$.
In Fig.\ref{fig1b}, ${\rm {\rm {C_{c}}}}$ is measured under different
ethanol concentrations at $T=20~^{\circ}{\rm {C}}$. Adding ethanol
to a solution effectively changes its dielectric constant. $\varepsilon$
for a mixture of ethanol and water is determined by $\varepsilon(T)=(1-c_{e})\varepsilon_{w}(T)+c_{e}\varepsilon_{e}(T)$~\cite{Neel_JPDAP1983},
where $\varepsilon_{w}(T)$ and $\varepsilon_{e}(T)$ are dielectric
constants of water and ethanol at temperature $T$ respectively, and
$c_{e}$ is the volume fraction of ethanol in the solution. At
20 $^{\circ}{\rm {C}}$, $\varepsilon_{w}=80.37$ and $\varepsilon_{e}=25.00$.
According to the relation between $\varepsilon$ and $c_{e}$, the
dielectric constant of a solution drops when its ethanol concentration
increases (Fig.\ref{fig1b}). The data in Fig.\ref{fig1b} show that
a decrease in $\varepsilon$ leads to decrease of ${\rm {C_{c}}}$.
Similar effect of ethanol on critical trivalent counterion concentration
has been observed in DNA condensation experiments~\cite{Bloomfield_BP1995}.
The relationship between ${\rm {C_{c}}}$ and $\varepsilon$ indicates
that the like-charge attraction originates from electrostatic correlations
between counterions. Since the strength of counterion correlations
is proportional to $1/\varepsilon$, 
a smaller $\varepsilon$ leads to a stronger counterion correlation, indicated by a lower ${\rm {C_{c}}}$.

\begin{figure}[f]
\includegraphics[width=0.5\linewidth]{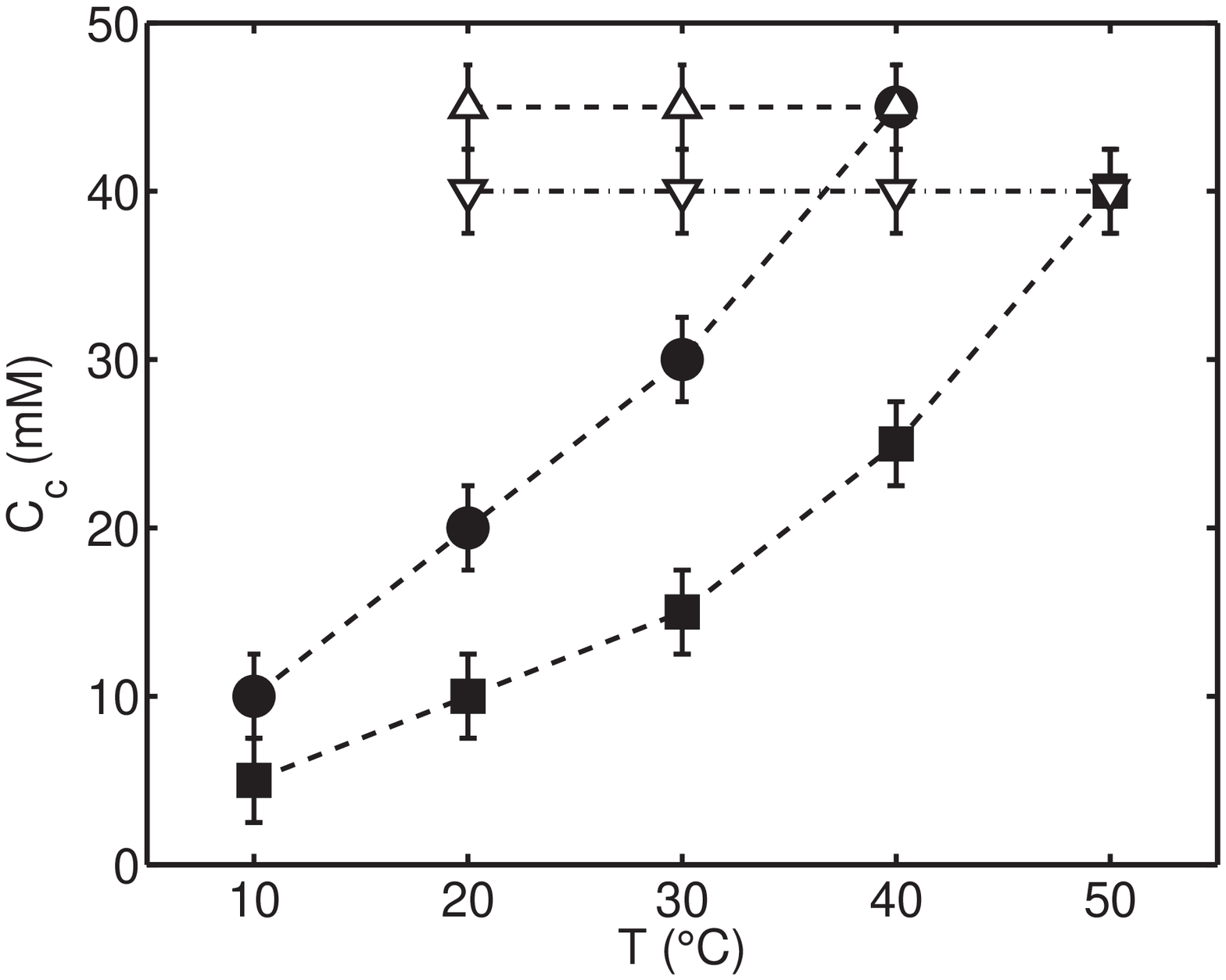} 
\caption{Effect of temperature on ${\rm {C_{c}}}$ under fixed dielectric
constants and fixed $l_{B}$. At $\varepsilon=73.2$ (circles)
and $\varepsilon=69.9$ (squares), ${\rm {C_{c}}}$ increases
with increasing temperature. When the solution dielectric constant
is adjusted to fix $l_{B}$, ${\rm {C_{c}}}$ doesn't change with
temperature . At fixed Bjerrum lengths, ${\rm {C_{c}}}$ is 45 mM
with $l_{B}=7.28$ (up triangles), and 40 mM with $l_{B}=7.38$ (down triangles).
\label{fig2}}
\end{figure}

Since ${\rm {C_{c}}}$ decreases upon the decrease of
$\varepsilon$, the dependence of ${\rm {C_{c}}}$ as a function of
temperature under a fixed dielectric constant is required to
manifest the pure temperature effect. In Fig.\ref{fig2}, ${\rm
{C_{c}}}$ increases as $T$ increases under fixed dielectric
constant. $\varepsilon$ of the sample is tuned to be constant
under different temperatures by the addition of appropriate
amounts of ethanol. At $\varepsilon=73.2$, ${\rm {C_{c}}}$ for
${\rm {MgCl_{2}}}$ increases from 10 mM to 70 mM as we raise the
temperature from $10~^{\circ}{\rm {C}}$ to $40~^{\circ}{\rm {C}}$.
Hence, the drop of ${\rm {C_{c}}}$ in Fig.\ref{fig1a} is actually
the net effect of an increase in temperature and a decrease of the
dielectric constant, which is induced by the rising temperature.
The curve for $\varepsilon=69.9$ is lower than that of
$\varepsilon=73.3$. This is to be expected, because lower
dielectric constants lead to lower threshold concentrations.

A decrease in either $\varepsilon$ or temperature leads to a lower
threshold concentration. We speculate that the effect of temperature
and dielectric constant can be combined to that of one parameter:
the Bjerrum length $l_{B}=e^{2}/4\pi\varepsilon\varepsilon_{0}k_{B}T$,
which is often applied to quantify the importance of counterion correlations
through a coupling parameter $\Xi=2\pi z^{3}l_{B}^{2}\sigma$, where
$z$ is the valence of counterions and $\sigma$ represents the surface
charge density of polyelectrolytes~\cite{Netz_PRL2001,Shklovskii_PRE1999}. 
At strong correlation limit, i.e. $\Xi\gg1$, the condensed counterions form 
two dimensional (2-D) strongly-correlated liquid or even Wigner crystals
at low temperatures.
${\rm {C_{c}}}$ is indeed a function of $l_{B}$, since it
doesn't change with temperature at a constant Bjerrum length (See Fig.\ref{fig2}). 
More importantly, a lower ${\rm {C_{c}}}$
is observed at a lower Bjerrum length. Thus, a stronger correlation
between counterions leads to a smaller critical counterion concentration,
since larger $l_{B}$ values indicate stronger counterion correlation.

\begin{figure}[f]
\includegraphics[width=0.5\linewidth]{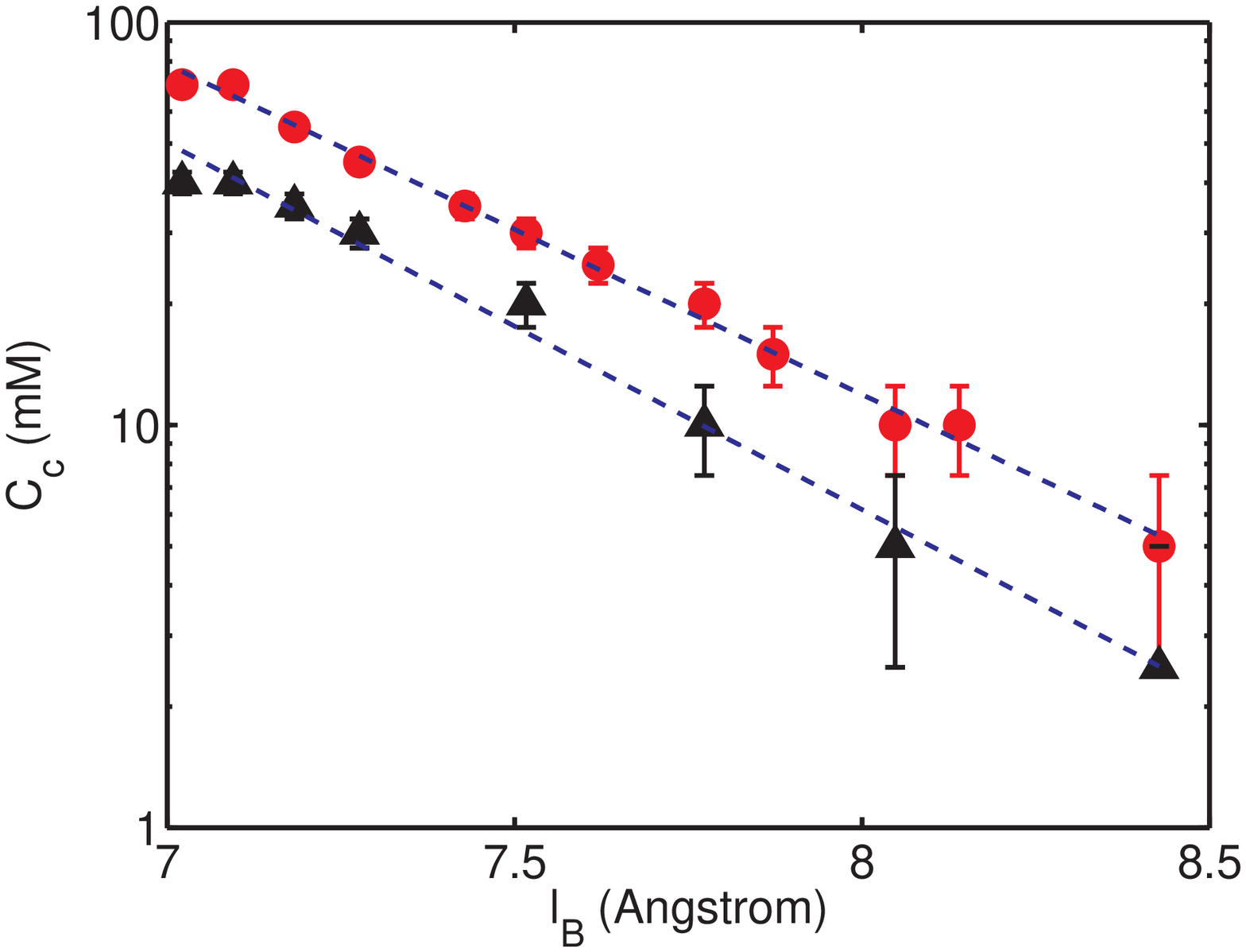}
\caption{${\rm {C_{c}}}$ as a function of $l_{B}$. Results for ${\rm {MgCl_{2}}}$
(filled triangles) and ${\rm {CaCl_{2}}}$ (filled circles) indicate
that ${\rm {C_{c}}}$ decreases exponentially with increasing $l_{B}$.
Dashed lines are results of exponential fittings to the experimental
results.\label{fig3}}
\end{figure}

The threshold concentration
is quite sensitive to the change in Bjerrum length. In Fig.\ref{fig3},
${\rm {C_{c}}}$ of ${\rm {MgCl_{2}}}$ and ${\rm
{CaCl_{2}}}$ decrease exponentially as a function of $l_B$.
The logarithm of ${\rm {C_{c}}}$ fits to a linear function of Bjerrum
length: $\ln{\rm {C_{c}}}=-Al_{B}+B$ , with $A=1.89~\text{\AA}^{-1}$, $B=17.6$
for ${\rm {MgCl_{2}}}$ and $A=2.09~\text{\AA}^{-1}$, $B=18.5$ for ${\rm
{CaCl_{2}}}$, respectively. The linear relationship between
$\ln{\rm {C_{c}}}$ and $l_{B}$ indicates that the counterion
entropy plays an important role in the process of aggregation. 
The difference in entropy for the counterions in a
bundle and those in the bulk medium is approximately $\ln{\rm
{C_{c}/C_{0}}}$, where ${\rm {C_{0}}}$ is the counterion
concentration in the condensation layer. This entropy loss of the
condensed counterions causes an increase in the energy barrier which
has to be overcome by the attractive interaction induced by
electrostatic correlations. Since the energy gain from
electrostatic interactions is proportional to $l_{B}$, a linear
relation between $\ln{\rm {C_{c}}}$ and Bjerrum length is
expected.

Based on the mean field theory approximation, the Oosawa theory predicts
that the force per unit length between two polyelectrolytes as~\cite{Oosawa_PE1968,Jensen_PRL1997}:
\[f(R)=k_{B}T(\frac{1}{z^{2}Rl_{B}}-\frac{z^{2}\xi^{2}}{(1+z^{2}\xi^{2})R^{2}})\]
where $z$ is the valence of counterions, $R$ is the distance
between the long axes of two parallel polyelectrolytes, and
$\xi=l_{B}/b$ is the Manning parameter. For highly charged
polyelectrolytes, $\xi\gg1$, $f(R)$ is attractive when the distance
$R$ between two charged lines is less than $z^{2}l_{B}$. To induce
aggregation of fd virus with divalent counterions ($z=2)$ at a
large $l_{B}$, say $l_{B}=9$ \AA, the distance between two
neighboring viruses must be smaller than 4 nm. However, the
distance between the long axes of two parallel viruses could not
be smaller than the diameter of a virus, which is approximately
6.6 nm. Therefore, it would be impossible for divalent counterions to
induce aggregations of fd virus solely based on this treatment. 
The Oosawa theory fails because of its mean field nature and the 
simplification of polyelectrolytes as charged lines.

The Wigner crystal model takes into account the strong correlation
between counterions. Before aggregation, each fd virus is
surrounded by a layer of condensed counterions, which partially
compensates the charge of the fd virus. As a result of the
decrease in fd surface charge, the two viruses can move close to
each other by Brownian motion, since the Coulomb repulsion
is reduced. When the distance between two viruses is small enough,
the counterions on different viruses start to correlate with each
other. This correlation creates an attractive interaction between
two viruses. Each counterion gains an energy of $G_{corr}$ from
this attractive interaction. Since $\Xi\approx14.8$ for fd virus at 
room temperature and $\varepsilon=80$, the condensed counterions can be treated as
a strongly correlated liquid, and $G_{corr}$ is then estimated as the 
energy of interaction between a counterion and its Wigner-Seitz 
cell~\cite{shklovskii_PRL99}.
\begin{equation}
G_{corr}=\frac{z}{2b}(zb/r)^{3/4}l_{B}k_{B}T
\label{eq1}
\end{equation}
 where $b$ is the charge spacing on the polyelectrolyte when considering
it as a charged line, and $r$ is the radius of a virus. $G_{corr}$
is estimated to be $0.13\phantom{.}l_{B}k_{B}T$ per counterion within a bundle
of fd viruses using $b=0.8$ \AA~ and $r=33$ \AA.

The strong correlation also drives more counterions from the bulk
solution into the space between two viruses~\cite{Nguyen_JCP2000,Grosberg_RMP2002}. These extra
counterions roughly neutralize the residual charge of fd viruses,
thus diminishing the Coulomb repulsion. However, the entropy loss
of these counterions incurs an energy penalty. Assume that a
fraction $\gamma$ of all counterions in the condensation layer are
the extra ones, the average energy penalty per counterion is
\begin{equation}
G_{R}=\gamma k_{B}T\ln{\rm {C_{0}/C}}
\label{eq2}
\end{equation}

Aggregation of fd viruses occurs when the energy gain is larger
than the energy penalty, $G_{corr}-G_{R}>0$. Using Eqs.(\ref{eq1})
and (\ref{eq2}), the critical concentration $\rm{C_c}$ can be
calculated as
\begin{equation}
\ln{\rm {C_{c}}}=-\frac{z^{7/4}b^{1/4}}{2\gamma r^{3/4}}l_B+\ln{\rm {C_{0}}}
\label{eq3}
\end{equation}

From Eq.(\ref{eq3}), $\gamma$ can be estimated by setting the fitting
parameter $A=\frac{z^{7/4}b^{1/4}}{2\gamma r^{3/4}}$. Based on the values of~$A=1.89~\text{\AA}^{-1}$~and $2.09~\text{\AA}^{-1}$, 
we obtain~$\gamma\approx6\%$~and 7\% for~$\rm{CaCl_2}$~and~$\rm{MgCl_2}$, respectively.  When
calculated from the Manning counterion condensation theory,
approximately $94\%$ of the surface charge of a virus is
neutralized by divalent counterions at the onset of aggregation.
These predictions suggest that the viruses are totally neutralized when they aggregate. 

The second term in Eq.(\ref{eq3}) contributes to the fitting
parameter $B$. The counterion concentration in the condensation
layer has been measured to be a few molars~\cite{Manning_QRB1978}.
Using ${\rm {C_{0}}}=3$ molar for both $\rm{Mg^{2+}}$ and
$\rm{Ca^{2+}}$, the parameter $B$ is estimated to be 8.01, which
is much smaller than the fitting result. The discrepancy likely
originates in the change in counterion hydration state. A
counterion loses its waters of hydration when it resides in the
condensation layer. The $^{25}\rm{Mg}$ NMR measurements of F-actin solutions 
indicate that only a small fraction of the condensed $\rm{Mg^{2+}}$ lose 
waters of hydration~\cite{xian_Biochem1999}. If a small
portion, say 10\%, of the extra counterions lose their waters of
hydration, an energy penalty of approximately $10k_BT$ is expected 
for each extra counterion, since the energy penalty for a $\rm{Mg^{2+}}$ or
$\rm{Ca^{2+}}$ ion to lose all of its waters of hydration is on
the order of $100k_BT$~\cite{pavlov_JPCA1998}. When taking into account
the hydration effect, $G_R=\gamma k_{B}T(\ln{\rm {C_{0}/C}}+10)$. 
Thus, $B=\ln{\rm{C_0}}+10\approx18$, which is close to the fitting results. This
estimate suggests that partial loss of hydration contributes to the change in Gibbs free
energy, thereby affecting the onset
concentration of multivalent counterions for bundle formation.

In the DNA condensation experiments~\cite{Parsegian_BJ1992}, the threshold
concentration of ${\rm {Mn^{2+}}}$ ions to induce DNA condensation
has also been reported to decrease with increasing temperature. The
phenomenon is similar to the result in Fig.\ref{fig1a}. When taking into account the variation in solution dielectric
constant, the data in Ref.~\cite{Parsegian_BJ1992} indicate an exponential
decay of ${\rm {C_{c}}}$ with increasing $l_{B}$ as well. Instead of discussing
the effect of variations in dielectric constant, the data was interpreted by 
assuming that the attractive force originates
in the release of ordered water from the polyelectrolyte surfaces~\cite{Parsegian_BJ1992,Rau_Biochem2004}. It is interesting that the analysis based on the water exclusion also predicts an extra binding of counterions in the process of DNA condensation \cite{Rau_Biochem2004}.

In summary, we have studied the thermodynamic effect of like-charge
attraction by measuring the temperature effects on threshold
concentrations for divalent counterions to induce bundle formation
of fd viruses. Increases in both the temperature and the dielectric 
constant of the solvent lead to lower critical counterion concentration. 
We found that ${\rm {C_{c}}}$ varies as a
function of Bjerrum length, which combines the effects of
temperature and the dielectric constant. The Oosawa model fails to predict the aggregations
of fd virus because it simplifies the polyelectrolyte rods to charged lines.
The linear relation between $\ln{\rm {C_{c}}}$ and $l_{B}$ is well interpreted using the
Wigner crystal model, supporting the argument that it is applicable at finite temperatures.

The authors appreciate the helpful discussions with Prof. B. I. Shklovskii,
Dr. Thomas Angelini, and Dr. Don Rau. The work is supported by NSF
DMR0405156 and NIH R01 HL67286.

\bibliography{fdtemp}

\end{document}